\documentclass[aps,prl,twocolumn,superscriptaddress,showpacs]{revtex4}

\usepackage[dvips]{graphics}
\usepackage{epsfig}
\usepackage{amsfonts}
\begin{document}
\newcommand{\beq}{\begin{equation}}
\newcommand{\eeq}{\end{equation}}
% Use the \preprint command to place your local institutional report
% number in the upper righthand corner of the title page in preprint mode.
% Multiple \preprint commands are allowed.
% Use the 'preprintnumbers' class option to override journal defaults
% to display numbers if necessary
%\preprint
% vVersion 1.1
%Title of paper
\title{Signatures of direct double ionization  under XUV radiation}
% repeat the \author .. \affiliation  etc. as needed
% \email, \thanks, \homepage, \altaffiliation all apply to the current
% author. Explanatory text should go in the []'s, actual e-mail
% address or url should go in the {}'s for \email and \homepage.
% Please use the appropriate macro foreach each type of information
% \affiliation command applies to all authors since the last
% \affiliation command. The \affiliation command should follow the
% other information
% \affiliation can be followed by \email, \homepage, \thanks as well.
\author{P. Lambropoulos} 
\affiliation{Department of Physics, University of Crete, P.O Box 2208, Herakleion 71003, Crete, Greece}
\affiliation{Institute of Electronic Structure and Laser, F.O.R.TH, P.O. Box
  1527, Herakleion 711 10, Crete, Greece}
\author{L. A. A.  Nikolopoulos} 
\affiliation{Department of Telecommunication Science and Technology, University of Peloponnese, Greece}
\author{M. G. Makris}
\affiliation{Department of Physics, University of Crete, P.O Box 2208, Herakleion 71003, Crete, Greece}
%\email[e-mail:]{makris@physics.uoc.gr}

%\homepage[]{Your web page}
%\thanks{3452}
%\altaffiliation{}
\date{{\bf version \today}}
% insert suggested PACS numbers in braces on next line

% insert suggested keywords - APS authors don't need to do this
%\keywords{}
%\maketitle must follow title, authors, abstract, \pacs, and \keywords
\begin{abstract}  
In anticipation of upcoming two-photon double ionization of atoms and particularly Helium, 
under strong short wavelength radiation sources (45 eV), we present quantitative signatures of direct 
two-photon double ejection, in the photoelectron spectrum (PES)  and the peak power dependence, 
that can be employed in the interpretation of related data. We show that the
PES provides the cleanest signature of the process. An inflection (knee) in
the laser power dependence of double ionization is also discernible, within a
window of intensities which depends on the pulse duration and cross sections 

%depending
%on the value of the relevant cross section and provided the intensity is above $10^{14}$ W/cm$^2$
%for a 30 fs pulse.
\end{abstract}  

\maketitle

% body of paper here - Use proper section commands
% References should be done using the \cite, \ref, and \label commands
%\section{}
% Put \label in argument of \section for cross-referencing
%\section{\label{}}
%\subsection{}
%\subsubsection{}

% If in two-column mode, this environment will change to single-column
% format so that long equations can be displayed. Use
% sparingly.
%\begin{widetext}
% put long equation here
%\end{widetext}

The issue of direct versus sequential double ionization has in the last few years emerged in a new context, 
namely two-photon double ionization of Helium under XUV radiation, and in particular photon energies 
of about 45 eV 
\cite{kornberg:1999,pindzola:1998,nikolopoulos:2001a,makris:2001,mercouris:2001,nikolopoulos:2002c,hart:2003}.
%\cite{pindzola:1998,kornberg:1999,nikolopoulos:2001a,parker:2001,makris:2001,mercouris:2001,nikolopoulos:2002c,hart:2003,piraux:2003,bachau:2003,colgan:2002a}. 
Although until very recently, sources of radiation in that wavelength range, 
mostly synchrotrons, could not provide the needed intensity (more than $10^{12}$ W/cm$^2$), 
the situation has now changed. It is conceivable that further developments and optimization of 
High Order Harmonic Generation (HOHG) might succeed \cite{watanabe:2004}. The upcoming second phase of the FEL XUV source at 
DESY \cite{ayvazyan:2002}, however, is expected to
easily satisfy that requirement. Thus it is a matter of probably short time
that the first experimental data on this process will be obtained. When that happens, it is important to have available unequivocal 
and quantitative signatures of the process and this is the purpose of this paper. 

What is it that makes this process interesting?  Recall that single-photon double ionization, 
especially in Helium, has been studied in great detail, both theoretically and experimentally 
\cite{byron:1967}.
%\cite{byron:1967,hino:1993,anderson:1993,samson:1994,pont:1995,tangj:1995,kheifets:1996,sadeghpour:1996,forrey:1997}.
It is basically well understood, although interesting details, especially near the threshold keep 
coming up \cite{e2e:2003}. It could also be argued that this process is fundamentally two-step, in the sense
that the single available photon can only interact with one of the electrons
and it is only through electron-electron correlation that double ejection is
possible. As often said, correlation either in the initial or the final state
is necessary \cite{dalgarno:1960}. To stress the point, let us note that the process would be
impossible for non-interacting electrons.  The same is true for the other
extreme case of double ionization, namely long-wavelength ($\sim$780 nm) high
intensity and short pulse duration. The mechanism for that process, other than
the sequential, is also explicitly understood as two-step 
\cite{Corkum:1993}.  Specifically, the
theoretical interpretation rests on the physical picture of one electron
pulled out by the strong oscillating field, set into oscillation and
liberating the other electron - with a probability depending on the intensity
- as it is driven back to the vicinity of the nucleus.  Despite the enormous
number of photons streaming through the atomic diameter, at those intensities,
these photons cannot act simultaneously and separately on each of the
electrons, with a probability of any significance.  The reason is screening,
even in a two-electron system, which does not allow the long-wavelength
photons to "see" both electrons "at the same time"; i.e. before one electron leaves and the other relaxes.
 Obviously, the same is true for atoms with more electrons, such as the rare gases, where double
ionization has been observed under such conditions and interpreted similarly.
This two-step process is referred to as non-sequential, to distinguish it
from the sequential double ionization, in which one electron is ionized by the
field  with the subsequent ionization of the ion, through a second high order process. 
The term direct has also been used for the non-sequential. We will, however, need to reserve 
here the term direct, for a somewhat different process, when it comes to the interaction
with photons of energy above  $\sim$40 eV and below 80 eV where the single-photon channel opens.

We shall be even more specific and consider the photon energy range $\sim$40-54 eV, with emphasis at $\sim$45 eV,
for reasons that we explain now.  Inspection of the energy level structure of Helium, with a first ionization 
threshold at $\sim$25 eV and the lowest doubly excited (autoionizing) state at $\sim$56 eV, shows that the
 absorption of a photon of 45 eV raises the system to a virtual state within the single electron continuum,
 detuned by $\sim$10 eV from the nearest discrete state. The absorption of a second photon, assuming sufficient 
intensity (i.e. flux of photons) can, among other things, lead to double ejection, through a "direct", 
and obviously non-sequential, process in which both electrons are acted upon by the photons independently. 
 The reason is that, because of the short wavelength, screening is of no significance. Moreover, this 
two-photon double ionization would occur even if the two electrons were non-interacting particles. It is 
in order to stress this feature, that we would propose to reserve the term direct for processes of this type, 
which are possible even in the absence of interaction, which entails correlation. In addition to the direct
 double ejection, producing He++, other processes that will take place, with the respective branching ratios,
 are: (a) Single-photon ionization producing He+(1s); by far the strongest
 channel, with a cross section $\sigma_a = 2.4 \times 10^{-18}$ cm$^2$. (b)
 Two-photon ionization of He+, producing He++, with a generalized cross section $\sigma_b = 1.0 \times10^{-53}$ cm$^4$ s  . 
(c) Two-photon ionization of He (ATI), leading to He+(1s) or even excited
 states. (d) An, in principle, infinite sequence of higher order processes,
 which owing to the combination of (short) wavelength and intensity of
 interest are of no significance in the context of this paper. The
 quantities $\sigma_a$ and   $\sigma_b$ can be calculated very accurately. The
 cross section $\sigma_2$ for the direct involves uncertainties as discussed below.

Let us now define the context.  We have in mind possible observations with sources such as those mentioned above. 
 To the best of our knowledge and estimate, one can not expect at present more than $10^{13}$ or possibly 
$10^{14}$ W/cm$^2$ from HOHG sources.  To be generous, however, let us say  $10^{16}$.
On the basis of what we know,  $10^{15}$ would be a hopeful intensity, at least for the initial operation of 
the next FEL phase.  The relevant pulse durations could range from a few tens of fs  (for HOHG) to around 100 fs 
for the FEL.  First note that, at $10^{16}$ W/cm$^2$, and photon energy 45 eV, the ponderomotive energy of the 
electron is $\sim$0.5 eV which is $1\%$ of the photon energy.  This means
 that ATI beyond the first peak  and related non-perturbative 
effects can be ignored. In addition, even  a pulse  duration of one fs is very much longer than one cycle 
($\sim$0.1 fs) of the field, which means that a transition rate is extremely well justified. Obviously, 
the above conditions also imply that recollision processes, which are crucial at 780 nm, are completely 
insignificant here.  Which is another way of saying that whatever double ionization is observed, will 
come either from the direct, as defined above, or the sequential. The transition probability per unit time 
obtained through lowest-order perturbation theory (LOPT) is therefore  meaningful, 
but it requires an accurate structure calculation for He, including the double continuum in the final state, 
for the direct process. The single active electron 
approximation, valid for long wavelength, is totally inappropriate here. In performing the summation 
over intermediate states, since the first photon reaches into the continuum, a pole is involved, 
which however can be handled (with care) using appropriate techniques  \cite{nikolopoulos:2001a,cormier}. 
For our purposes here, and the needs of interpretation of experimental data, in this context, 
it is mainly processes (a) and (b), in addition to the direct, that are of relevance. 
The role of processes (c) is discussed later on. It is important to note that, for photon energies around 45 eV, 
process (b) requires two photons, which makes the sequential of third order, compared to the second order 
of the direct \cite{kornberg:1999}. This feature, favoring the direct quantitatively and spectroscopically 
as we shall see below, is lost above $\sim$ 54 eV. Obviously, when the
 intensities reach values considerably higher, time-dependent solutions of the
 Schr\"odinger equation, such as those already presented in references 
\cite{pindzola:1998}, among others, must come into play.

This is not the place to elaborate on the relevant theoretical techniques. It should suffice to say that 
the transition rate for the direct process has been calculated by a number of authors 
\cite{kornberg:1999,pindzola:1998,nikolopoulos:2001a,mercouris:2001,hart:2003}.
Strictly speaking, in some cases, it is the ionization yield, in a time dependent approach, that has been reported, 
from which a generalized two-photon cross section can be readily extracted, since the conditions of validity 
are satisfied.  
Although the results have not yet reached the degree of accuracy found in single-photon double ionization, 
it can be reasonably said that the cross section is known well to within one order of magnitude.  Actually, 
most calculations agree to within a factor of 2, but we shall try to be on the safe side. 
%%%%%%%%%%%%%%%%%%%%%%%%%%%%%%%%%%%%%%%%%%%%%%%%%%%%%  fig1
\begin{figure}
\centerline{\hbox{\psfig{figure=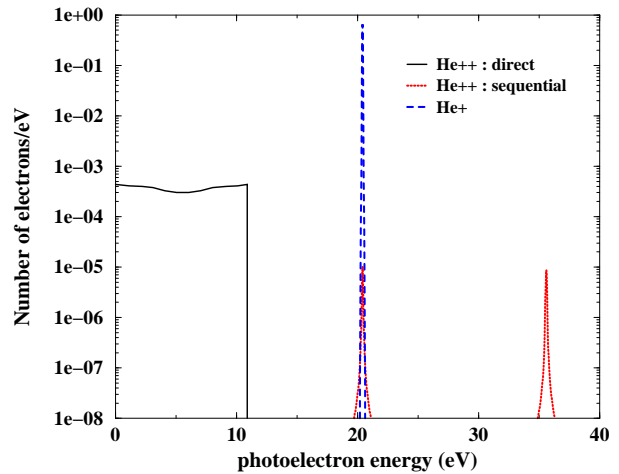,width=8cm}}}
\caption{Two-photon PES of Helium under the irradiation with a Gaussian pulse of photon energy 45 eV, 
peak intensity \protect 10$^{14}$ W/cm$^2$ and 30 fs duration. Only the
ionization paths (a) and (b) have been considered, in addition to the direct,
with $\sigma_2=8.1 \times 10^{-52}$ cm$^4$ s.}
\label{fig1}
\end{figure}
%%%%%%%%%%%%%%%%%%%%%%%%%%%%%%%%%%%%%%%%%%%%%%%%%%%%%  fig1

Perhaps the cleanest signature of the direct process is to be found in the
angle-integrated (PES). Its general outline was  presented in \cite{kornberg:1999} together with an  estimate of the relevant cross sections.  
What is needed now is a quantitative picture, based on up to date information on the cross sections and 
laser parameters expected to be available.  
Thus, employing the approach  in
\cite{nikolopoulos:2001a} and a rather generous value for $\sigma_2$, we present in Fig. \ref{fig1}, an example of the PES calculated under the conditions of intensity and pulse duration given in the caption.  Note that the figure provides a quantitative result for the differential 
ion yield. The yield will scale according to the power of the intensity for the respective 
process and proportionally to the duration. In addition, the widths of the narrow peaks will change somewhat.
  The structure of the PES is due to the direct process plus (a) and (b) and will not change for different 
combinations of intensity and pulse duration. In particular, the relative position of the peaks, 
and the energy separation of the direct from the dominant peak (a) will not change appreciably. 
This is what makes this double ionization process very 
special.  Obviously, all electrons with kinetic energy less than $\sim$11 eV, originate from the direct.   
Initial but somewhat limited theoretical studies \cite{makris:2001} have given hints of rather unexpected 
behaviour of the photoelectron angular distribution and its connection to correlation, as a function of 
the partition of the excess energy between the two ejected electrons. These would surely provide further 
signatures, although much more demanding experimentally. For the moment, it is safer to focus on the 
requirements for the detection and analysis of the angle-integrated signal. Perhaps the most serious challenge 
to a measurement of the spectrum  comes from the high but narrow peak at $\sim$20 eV, which, although well 
separated energetically,  could mask the signal 
from the direct. It should in addition be noted that processes (c) would produce a series of peaks, 
between the edge of the direct and (a), but of much smaller and diminishing height. We understand that 
electron-electron coincidence measurements \cite{private} may be necessary to cope with that  "noise"; 
an issue  well beyond our expertise. On the basis of the approach outlined here, one can readily 
calculate spectra for a variety of experimental conditions, which would, however, be meaningful only 
in relation to the specifics of the contemplated experiment.

Turning now to a second possible signature, we consider the ion yield, for both He+ and He++, as a function 
of the laser power.  This has in fact  been the basic tool in the identification and study of non-sequential 
double ionization in the long-wavelength regime \cite{fittinghoff:1992}.  
Can it also serve equally well in this case of a very different non-sequential process? 
Needless to say, detecting the ions is considerably less demanding experimentally. 
 The question is whether it can provide sufficient information to at least identify the presence 
of the direct process.  The direct does of course contribute at any intensity and pulse duration, 
with a branching  ratio determined by the atomic and source parameters.  The former being fixed, 
it is the latter that will decide whether its contribution may be detectable or hopelessly beyond reach.  

To evaluate its relative importance,  we have considered the set of 
differential equations \cite{tang:1987} governing the rate of production and destruction 
of the three species, namely He, He+ and He++ under a pulsed source of prescribed parameters.
Let  $N_0, N_1, N_2$ be the number of He, He+ and He++ in the interaction volume. 
Their evolution during the pulse obeys the equations:
\begin{eqnarray}
\dot{N}_0   &=& -\sigma_a F(t) N_0    -  \sigma_2 F^2(t) N_0,         \nonumber \\
\dot{N}_1   &=& \sigma_a  F(t) N_0   -  \sigma_b F^2(t) N_1, \nonumber \\
\dot{N}_{2,dir}   &=& \sigma_2 F^2(t) N_0,                            \nonumber \\
\dot{N}_{2,seq}   &=& \sigma_b F^2(t) N_1,                  \nonumber 
\end{eqnarray}
with $N_2={N}_{2,dir}+{N}_{2,seq}$ and F(t) is the photon flux. 
%The cross sections employed are 
%$\sigma_1 = 2.4 \times 10^{-18}$ cm$^2$ for (He) + $\hbar \omega \to$ He+, 
%$\sigma_2 = 1.0 \times10^{-53}$ cm$^4$ s  for (He+) + $2\hbar \omega \to$ He++, 
%and  $\sigma_2 = 1 \times 10^{-52}$ cm$^4$ s  for (He)  + $2\hbar \omega \to$ He++.
%%%%%%%%%%%%%%%%%%%%%%%%%%%%%%%%%%%%%%%%%% fig2
\begin{figure}[!t]
\centerline{\hbox{\psfig{figure=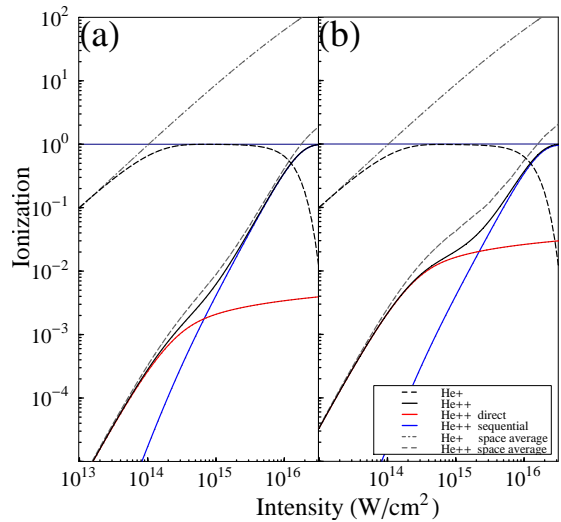,width=8cm}}}
\caption{ He$+$ and He$++$ yield obtained from a Gaussian laser pulse of 30 fs 
(full-width at half-maximum). (a) $\sigma_2=1 \times 10^{-52}$ cm$^4$ s 
(b) $\sigma_2=8.1 \times 10^{-52}$ cm$^4$ s.}
\label{fig2}
\end{figure}
%%%%%%%%%%%%%%%%%%%%%%%%%%%%%%%%%%%%%%%%%% 
Typical results about the expected dependence of the respective yields, in log-log plots, 
for the set of parameters indicated in the captions, are shown in Figs. 2a and 2b. In Fig. 2a, 
the value of the cross-section for the direct has been chosen such as to agree with the majority 
of the calculations published thus far, although one calculation
\cite{hart:2003} has given a somewhat lower value. It is evident that for peak intensity below about 
$10^{13}$ W/cm$^2$, double ionization is several orders of magnitude smaller than single ionization.  
Once double ionization begins making a relatively significant contribution, say above $ 10^{13}$, 
it is dominated by the direct, by several orders of magnitude. In fact up to $10^{14}$, the sequential 
double ionization is practically insignificant, becoming a non-negligible part
of double ionization, eventually taking over, arround $\sim 10^{15}$;
depending of course on the value of $\sigma_2$. 
In Fig. 2b, the value for the cross section of the direct has been chosen
larger by a factor of 8.  That is because an early calculation by two of us 
\cite{nikolopoulos:2001a} had given such an optimistic value, which turned out to disagree 
with calculations by others that followed, including some by us \cite{lambropoulos:e2e}. 
Nevertheless, the results of Fig. 2b are shown here as perhaps an upper bound. As could have been expected, 
the overall behaviour is similar to that of Fig. 2a, but shifted to lower intensities  The pulse duration of 
30 fs chosen for the calculations is somewhere in the middle of the range of durations expected for HOHG sources, 
on the one hand, and FEL on the other. Changing that value up to say 100-150 fs, or down to $\sim$10 fs, 
would not change the overall behaviour much. However, the contribution of the
direct will, at a certain peak intensity, begin increasing as the pulse
duration becomes shorter; i.e., when the saturation of He+ does not have the time
to drain the neutral species. If we were to summarize the message 
of these two figures, it appears that an intensity of at least $10^{13}$ W/cm$^2$, 
and preferably considerably more, is necessary for a relatively significant presence of the direct, 
and that more than $10^{15}$ or $5\times 10^{15}$, dependingon the cross sections, would hinder its observation. It is conceivable that our reading 
of the message of these two figures may seem incomplete to an experimentalist. In any case, here 
they are for the information and use of those interested.

An additional aspect in such plots, that has been fairly important, is the
spatial distribution of the intensity in the interaction volume. That is
because, due to the focusing, as the peak intensity rises, an increasing part
of the atomic beam begins  contributing to the ionization species. A
quantitative assessment of that effect requires the specifics of the spatial
distribution of the radiation for a given experiment.  Experience has shown,
however, that a typical form \cite{perry:1988} of the distribution provides useful insight into
the overall effect. We have performed such a sample calculation  for the set of parameters employed 
in Figs. 2a,b, with the result plotted on the same figure. As expected, there is no change 
for lower intensities (up to $\sim$$10^{14}$). Beyond that, the amounts of both He+ and He++ 
continue growing, as larger portions of the atomic beam produce significant signal. In all of the figures, 
with or without spatial integration, the curve for He++ exhibits an inflection (referred to as a "knee" 
in experiments at long wavelength \cite{fittinghoff:1992}). Its presence is a signature of non-sequential 
(whether from recollision or direct) double ionization, while the degree of its prominence is here 
seen to depend on the relative magnitude of the cross sections. Unlike the long wavelength case, 
where ab initio yields are practically impossible to come by, here we have a quantitative picture 
which, given sufficient information about the experimental parameters, can be directly related 
to the cross sections.

%In summary, we have given a number of quantitative signatures for direct double ionization under 
%strong XUV radiation which can serve as tools in the detection and study of direct processes. 
%Further more specific and detailed calculations must await the first
%experimental results. Nevertheless, certain aspects of our results are
%expected to be of general validity.  One is the shape and overall structure of
%the PES.  Another is the character of the direct and the window of
%intensities, in combination with the pulse duration, in which it makes a
%contribution sufficiently significant to separate it from the other competing
%processes.  This has been possible to identify here, in contrast to the long
%wavelength case, because ab initio values for the cross sections can  be
%calculated.           
% END of 4 pages
Although we have chosen a specific photon energy, as indicated
earlier, the overall behaviour is not expected to change much for photon
energies up to about 54 eV.  Finally, our particular choice of atom and order
of the process should not be interpreted as a unique context for the study of
these aspects of double ionization.  In principle, they can be sought in a
variety of atoms, in the appropriate range of wavelengths, one such example
having been discussed in Ref.\cite{Nikolopoulos:2003}.

We would like to acknowledge support from the European COST action D26/0002/02.
 One of us (PL) wishes to acknowledge discussions with H. Bachau and B. Piraux.
%\acknowledgments
\bibliographystyle{apsrev}
%\bibliography{/home/nlambros/mypapers/bib/Thesis}
%\bibliography{/home/makris/Documents/Bibliography.bib}

\end{document}